\documentclass[reprint, superscriptaddress, secnumarabic, amssymb, nobibnotes, aps, pra]{revtex4-1}

\usepackage{graphicx}
\usepackage{subfiles}
\usepackage{epstopdf}
\usepackage{parskip}
\usepackage[T1]{fontenc}
\usepackage[utf8]{inputenc}
\usepackage{newunicodechar}
\newunicodechar{⁢}{}
\newunicodechar{−}{\ensuremath{-}}
\usepackage{amsbsy}
\usepackage{gensymb}

\usepackage[T1]{fontenc}
\usepackage{amsmath}
\usepackage{amssymb}
\usepackage{bbm}
\usepackage{physics}
\usepackage{braket}
\usepackage{xcolor}
\usepackage{ wasysym }
\usepackage{float}
\allowdisplaybreaks
\usepackage{graphicx}
\usepackage[colorlinks=true]{hyperref}  
\hypersetup{
    bookmarks=true,         
    unicode=false,          
    pdftoolbar=true,        
    pdfmenubar=true,        
    pdffitwindow=false,     
    pdfstartview={FitH},    
    pdftitle={Ce2Rh3Ga5: A heavy fermion system with unusual magnetoresistance},    
    pdfauthor={Arushi et al.},     
    pdfsubject={},   
    pdfcreator={},   
    pdfproducer={}, 
    pdfkeywords={} {} {}, 
    pdfnewwindow=true,      
    colorlinks=true,       
    linkcolor=blue, 
    citecolor=blue,        
    filecolor=magenta,      
    urlcolor=blue           
} 
\usepackage[normalem]{ulem}

\newcommand{\figref}[1]{Fig.~\ref{#1}}

\renewcommand{\approx}{\simeq}

\makeatletter
\def\maketitle{
\@author@finish
\title@column\titleblock@produce
\suppressfloats[t]}
\makeatother

\begin{document}
\title{Anomalous Field-induced Coherence beyond Single-ion Kondo Physics}
\author{Arushi}
\email[al270@rice.edu]{}
\affiliation{Max-Planck Institute for Chemical Physics of Solids, 01187 Dresden, Germany}
\author{Konstantin Semeniuk}
\affiliation{Max-Planck Institute for Chemical Physics of Solids, 01187 Dresden, Germany}
\author{Manuel Brando}
\affiliation{Max-Planck Institute for Chemical Physics of Solids, 01187 Dresden, Germany}
\author{Javier F. Landaeta}
\affiliation{Max-Planck Institute for Chemical Physics of Solids, 01187 Dresden, Germany}
\author{Elena Hassinger}
\affiliation{Max-Planck Institute for Chemical Physics of Solids, 01187 Dresden, Germany}
\author{Christoph Geibel}
\email[Christoph.Geibel@cpfs.mpg.de]{}
\affiliation{Max-Planck Institute for Chemical Physics of Solids, 01187 Dresden, Germany}
\date{\today}
\begin{abstract}
\begin{flushleft}
\end{flushleft}
The emergence of coherent heavy quasiparticles in Kondo lattices is generally accompanied by a reduction in electrical resistivity at low temperatures. However, certain strongly correlated metals exhibit anomalously large resistivities that persist to very low temperatures, raising fundamental questions about the mechanisms governing charge transport. Here, we report an unusual magnetic field-induced evolution of electronic transport and thermodynamic properties in a Ce-based heavy fermion compound, Ce$_{2}$Rh$_{3}$Ga$_{5}$. At zero field, Ce$_{2}$Rh$_{3}$Ga$_{5}$ undergoes a magnetic transition at $T$$_{0}$ = 0.45 K and exhibits a strongly enhanced low temperature specific heat coefficient, reaching $C$/$T$ = 2.2 J/mol$_{Ce}$-K$^{2}$, but no discernible coherence in its electrical resistivity. Remarkably, applying a magnetic field produces a large negative magnetoresistance and restores coherent metallic behavior, characterized by low-temperature $T$$^2$ resistivity. Although the field dependence of the electronic specific heat coefficient can be described phenomenologically within a single-ion Kondo framework, the same description fails to account for the strong suppression of the residual resistivity and the overall magnetoresistance. These results point to an additional field-sensitive scattering contribution beyond conventional single-ion Kondo physics. The striking resemblance to the long-standing anomalous magnetotransport behavior of UBe$_{13}$, without the complications introduced by superconductivity, establishes Ce$_{2}$Rh$_{3}$Ga$_{5}$ as a platform for investigating the microscopic origin of anomalous scattering and field-induced coherent transport in strongly correlated electron systems.

\end{abstract}
\maketitle

\section{Introduction}
Electrical transport in strongly correlated electron systems is governed by the intricate interplay between localized magnetic moments and itinerant charge carriers \cite{Localizedmomnets_Heavyquasi_1, Localizedmomnets_Heavyquasi_2}. In heavy-fermion compounds based on Ce, Yb, and U, the Kondo interaction produces strong electronic correlations that substantially renormalize the quasiparticle masses and influence their scattering processes \cite{Localizedmomnets_Heavyquasi_3, Localizedmomnets_Heavyquasi_4}. At elevated temperatures, incoherent scattering from local magnetic moments dominates the resistivity, whereas lowering the temperature typically leads to the emergence of a coherent Kondo-lattice state, resulting in a typical resistivity maximum at intermediate temperatures, followed by a pronounced decrease at low temperatures \cite{Incoherent_Kondolattice_1, Incoherent_Kondolattice_2, Incoherent_Kondolattice_3}. Applied magnetic fields further modify these processes by suppressing spin-flip scattering and altering the characteristic energy scales of the correlated electronic state \cite{Hdependence_onKondo_1, Hdependence_onKondo_2, Hdependence_onKondo_3}. Although such behavior is well understood within the Kondo and heavy-fermion frameworks, the occurrence of exceptionally large low-temperature resistivities and their dramatic suppression by magnetic fields in certain strongly correlated materials remains a challenge \cite{Anomalous_transport_1}. Whether these anomalous transport responses can be explained entirely by conventional Kondo scattering or require additional field-sensitive scattering processes is an important unresolved question.

The heavy-fermion superconductor UBe$_{13}$ provides a particularly intriguing example of such anomalous transport. Unlike many conventional Kondo-lattice systems, UBe$_{13}$ enters its superconducting state at (T$_{c}$  = 0.9) K from a highly resistive normal state, with a broad resistivity maximum occurring only a few Kelvins above $T$$_{c}$ \cite{UBe13_1, UBe13_2, UBe13_3}. Applying a magnetic field produces an unusually large negative magnetoresistance and substantially modifies the low-temperature resistivity, including the emergence of a $T$$^{2}$ dependence in appropriate field and temperature regimes \cite{UBe13_3, UBe13_4, UBe13_5}. Remarkably, the residual resistivity inferred from the low-temperature normal-state transport is strongly field-dependent, suggesting that its large magnitude cannot be attributed solely to conventional, field-independent impurity scattering. Previous investigations have considered several possible origins of this behavior, including unconventional Kondo physics \cite{UBe13_6, UBe13_7}, intersite magnetic correlations \cite{UBe13_8, UBe13_9}, and proximity to quantum criticality \cite{UBe13_2, UBe13_5}. Nevertheless, a unified microscopic explanation of its anomalous low-temperature magnetotransport remains unresolved \cite{UBe13_9, UBe13_10}. One problem is the presence of superconductivity, which restricts direct access to the normal state at low temperatures and low magnetic fields, complicating efforts to disentangle competing scattering mechanisms.

\begin{figure*}
\includegraphics[width=2.0\columnwidth,origin=b]{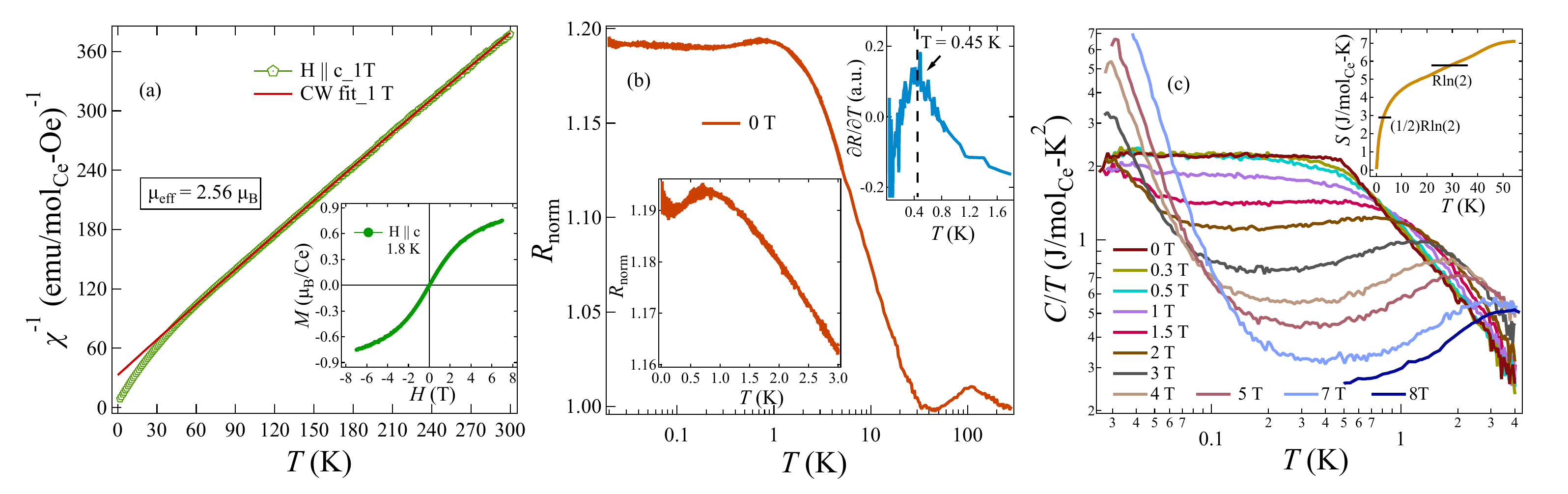}
\caption{\label{Fig1:ALL} a) Temperature dependence of the inverse susceptibility under a 1 T applied field for $H$ || $c$ with a modified Curie-Weiss fit in the high-T part. The inset shows the the field dependence of the magnetization at $T$ = 1.8 K for H || $c$. b) Electrical resistance w.r.t. temperature (log scale) in the whole investigated temperature range under zero applied field. The inset in the bottom left shows a zoomed view of low-temperature resistance, whereas the top-right inset presents the temperature dependence of the slope d$R$/d$T$, which reveals an anomaly at about 0.45 K, indicating some ordering. c) Temperature dependence of $C$/$T$ in zero and applied fields up to 8 T in a log-log plot. The inset shows the temperature evolution of entropy from 0.02 K $\leq$ $T$ $\leq$ 55 K.}
\end{figure*}

Here, we investigate a newly discovered Ce-based heavy-fermion compound, Ce$_{2}$Rh$_{3}$Ga$_{5}$, which crystallizes in the U$_{2}$Co$_{3}$Si$_{5}$-type Orthorhombic structure and exhibits striking similarities to the anomalous transport characteristics of UBe$_{13}$. Low temperature ac susceptibility, electrical resistivity, and specific heat measurements reveal a magnetic transition at $T$$_{0}$ = 0.45 K. Between 4 K and $T$$_{0}$, $C$/$T$ increases by nearly an order of magnitude and approaches a remarkably large, nearly temperature independent value of approximately 2.2 J/mol$_{Ce}$-K$^{2}$ below the transition. On the other hand, the zero-field electrical resistivity retains a predominantly incoherent Kondo-like temperature dependence, without a discernible crossover to conventional coherent Kondo-lattice transport. Remarkably, applying a magnetic field dramatically changes this behavior, producing a large negative magnetoresistance and restoring a low-temperature metallic state exhibiting Fermi-liquid-like transport. Unlike UBe$_{13}$, Ce$_{2}$Rh$_{3}$Ga$_{5}$ does not exhibit superconductivity over the investigated temperature range, permitting direct investigation of its low-temperature transport and thermodynamic properties across the entire accessible magnetic field range.

Our combined analysis of electrical resistivity and specific heat reveals an important distinction between the field evolution of quasiparticle thermodynamics and charge transport. Applying a magnetic field induced a well-defined Fermi-liquid behavior with a T-independent $C$(T)/$T$ = $\gamma$(H), $\rho$(T) = $\rho_{0}$ + $A$(H)$T$$^{2}$, and $A$(H) scaling with $\gamma^2$(H). Interestingly, the suppression of $\gamma$(H) can be described phenomenologically using a single-ion Kondo model. However, the same framework does not adequately reproduce neither the strong-field dependence of the residual resistivity nor the full temperature- and field-dependence of the magnetoresistance. Together with the absence of a simple universal magnetoresistance scaling, these observations suggest that the field-dependent charge transport involves additional scattering physics beyond the conventional single-ion Kondo response. The microscopic origin of this contribution remains to be established, with quantum interference effects as one possible scenario. The close correspondence between Ce$_{2}$Rh$_{3}$Ga$_{5}$ and UBe$_{13}$, despite their different $f$-electron configurations and low-temperature phases, suggests that these anomalous transport characteristics may reflect a more general feature of strongly correlated electron systems. Ce$_{2}$Rh$_{3}$Ga$_{5}$, therefore, provides an opportunity to investigate the interplay between Kondo correlations, electronic coherence, and unconventional low-temperature scattering in the absence of superconductivity.

\section{Results and Discussion}

\subsection{Heavy-Fermion Behavior}

The temperature dependence of the basic properties, magnetic susceptibility $\chi$(T), electrical resistance R(T), and specific heat $C$(T) indicates Ce$_{2}$Rh$_{3}$Ga$_{5}$ to be a system with trivalent Ce atoms with localized 4$f$ moments experiencing some hybridization with conduction electrons. This results in Kondo behavior that evolves into a heavy-fermion state at low Temperatures. Figure \ref{Fig1:ALL}(a) shows the T dependence of $\chi^{-1}$(T) of Ce$_{2}$Rh$_{3}$Ga$_{5}$ measured on a single crystal with the field along the [001] direction. In the $T$ range 90 K $\leq$ $T$ $\leq$ 300 K, $\chi^{-1}$(T) exhibits a linear $T$ dependence, as expected for a local moment system. This part is well described by the modified Curie-Weiss (CW) equation: $\chi^{-1}(T) = \left(\chi_{0} + {C}/{(T-\theta_{p}})\right)^{-1}$, where $\chi_{0}$ represents the temperature-independent Pauli susceptibility, $C$ is the Curie constant, and $\theta_{p}$ is the paramagnetic Curie temperature. The fit yields $\chi_{0}$ = 1.17(1)$\cross10^{-4}$ emu/mol$_{Ce}$, $C$ = 0.824(1) emu-K/mol$_{Ce}$, and $\theta_{p}$ = -27.23(1) K. The calculated effective magnetic moment ($\mu_{eff}$ = 2.56$\mu_{B}$/Ce) closely matches the expected theoretical value of 2.54$\mu_{B}$ for a Ce$^{3+}$ free ion, indicating well-localized Ce moments at high temperatures. The deviation from the CW fit below 90 K can be attributed to the crystal electric field (CEF).  In trivalent Ce systems with non-cubic local symmetry, the six-fold degenerate localized $J$ = 5/2 4$f$ state is split by the CEF into three doublets, with a splitting energy of the order of typically 200 K - 400 K \cite{CEF1, CEF2}. Therefore, while at high temperature the full $J$ = 5/2 multiplet is active, at low temperatures only the ground CEF doublet is relevant. The inset of \figref{Fig1:ALL}(a) presents $M$(H) isotherms collected at 1.8 K. The magnetization shows evidence for saturation to values of about 0.8 $\mu_{B}$/Ce for fields above 4 T. The saturation field is of the order expected for a free $S$ = $\frac{1}{2}$ ion at $T$ = 1.8 K, $\mu_{0}$H$_{sat}$ = k$_{B}$T/m$_{sat}$ $\approx$ 3.3 T, indicating the sum of exchange energies to be rather small. The saturation moment is slightly below the typical value expected for the CEF ground state in Ce systems \cite{NRef1}. However, a further increase in $M$(H) for $\mu_{0}$H > 4 T suggests some screening of the moment at low field due to the Kondo effect.

The presence of a Kondo interaction is evidenced by the T-dependence of $R$(T), which differs markedly from that of standard metallic systems. Because of the difficulty in evaluating the geometrical factor of the small, irregular crystals, $R$(T) is presented instead of $\rho$(T) in \figref{Fig1:ALL}(b). Below 300 K, instead of decreasing with T as expected for a standard metallic system, $R$(T) first increases, before presenting a maximum at about 100 K. This feature is attributed to the inelastic Kondo scattering from the excited CEF states \cite{CEF3, CEF4}, indicating that the energy splitting of the first excited CEF state is about 200 - 300 K. Below 100 K, $R$(T) decreases again and exhibits a minimum at T $\approx$ 45 K, resembling a Kondo minimum. From this minimum, $R$(T) increases logarithmically with decreasing $T$ down to 2 K, as expected for incoherent Kondo-type scattering. Below 2 K, rather than showing the strong decrease typical of Kondo lattices, $R$(T) exhibits only a very weak maximum at about 1 K, then levels off at a high residual value, $\rho_{0}$. The absence of the pronounced decrease expected in a Kondo lattice due to the formation of a coherent state indicates persistent strong scattering or fluctuations down to the lowest temperatures. This feature will be addressed later. Furthermore, a precise analysis of the data reveals a faint anomaly in $R$(T), best seen in the first derivative (upper inset of \figref{Fig1:ALL}(b)). This provides a preliminary indication of a phase transition.

Further information about the Kondo interaction can be obtained from the specific heat. In the main part of \figref{Fig1:ALL}(c), we show the specific heat of Ce$_{2}$Rh$_{3}$Ga$_{5}$ in a log-log plot of $C$/$T$ versus $T$ in the range 30 mK < $T$ < 4 K. The zero-field data show that below 10 K, $C$(T)/$T$ increases with decreasing temperature, down to 0.5 K, where it reaches a huge value of more than 2 J/mol$_{Ce}$/K$^{2}$. This increase is not as steep as expected and is observed just above the magnetic ordering temperature in standard magnetic materials. Together with the huge values observed below 0.5 K, this indicates Kondo-type behavior and/or strong fluctuations. Indeed, in the range 3 K < $T$ < 0.5 K, $C$/$T$ follows quite nicely a power law $C$/$T$ $\approx$ $T$$^{-0.9}$, suggesting some quantum critical fluctuations \cite{QCP_scaling}. However, at $T$$_{0}$ = 0.5 K, there is a sharp discontinuity in the slope d($C$/$T$)/dT, with $C$/$T$ staying constant at lower temperatures. This discontinuity in the slope is a strong indication of a collective transition. This shall be discussed in detail further below. The inset of \figref{Fig1:ALL}(c) illustrates the estimated magnetic entropy obtained by integrating $C$$_{mag}$/$T$ in the overall measured temperature range up to 55 K (shown in the supplementary section 4). Initially, the entropy increases rapidly up to 3 K, then more gradually, showing a tendency to saturate above 10 K at a value close to $R$ln(2)/Ce, the value expected for the doublet CEF ground state. This indicates that the CEF ground state doublet is well separated from the first excited doublet, with a splitting larger than 100 K. In such a case, an estimate for the strength of the interactions acting on the CEF ground state doublet is provided by taking twice the temperature at which half of the $R$ln(2) entropy is recovered \cite{entropy1, entropy2}. For Ce$_{2}$Rh$_{3}$Ga$_{5}$, ($\frac{1}{2}$)$R$ln(2) is reached at 2.7 K; thus, the overall energy scale of these interactions is of the order of 5 K. The absence of a strong magnetic transition, together with the T-dependence of $R$(T), indicates that the dominant interaction is of Kondo type. All the results presented up to now indicate Ce$_{2}$Rh$_{3}$Ga$_{5}$ to be a local moment system with a dominant Kondo interaction of the order of $T$$_{K}$ $\approx$ 5 K, resulting in a heavy fermion system at lowest $T$.

\begin{figure*}
\includegraphics[width=2.0\columnwidth]{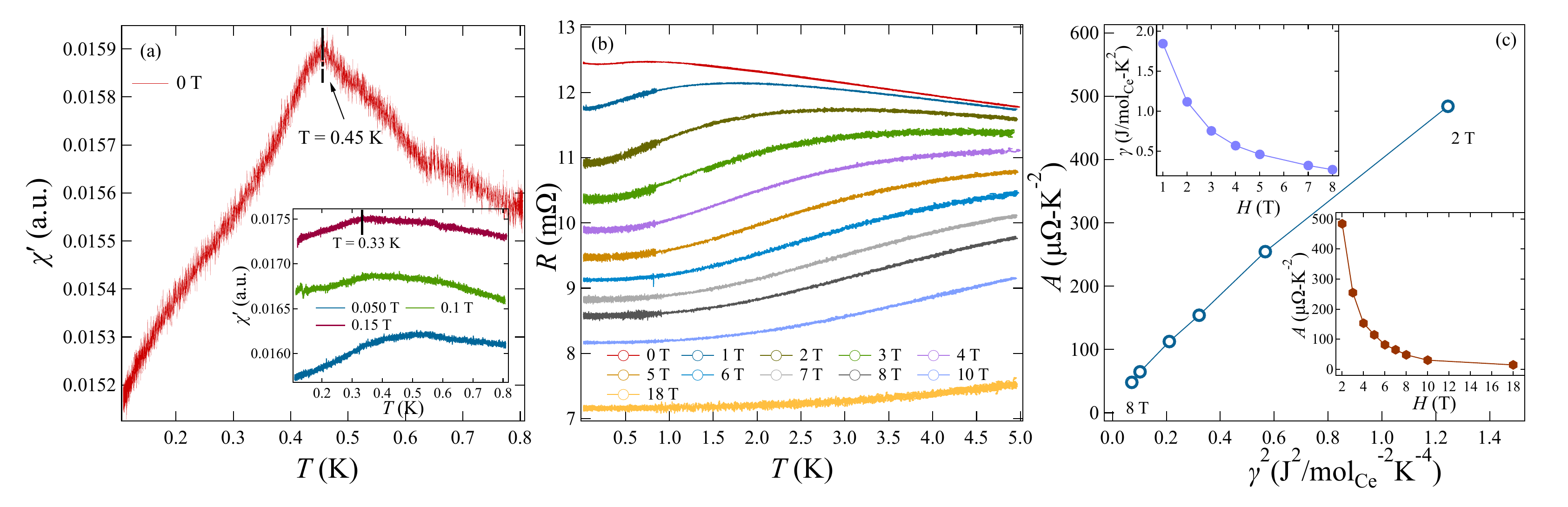} 
\caption{\label{Fig2:res} a) Temperature dependence of AC susceptibility measured in zero fields exhibiting an antiferromagnetic transition at 0.45 K. Inset shows the evolution of the magnetic ordering with increasing applied fields up to 0.15 T. b) $R$(T) in the low temperature range 0.02 K $\leq$ $T$ $\leq$ 5 K from 0 T to 18 T. c) The $T$$^{2}$ coefficient $A$ versus $\gamma^{2}$ from 2 T to 8 T following Kadowaki-Woods relation. The top-left and bottom-right insets show the field dependence of $\gamma$ and $A$, respectively.}
\end{figure*}

\subsection{Magnetic Order}

The sharp discontinuity in the slope d($C$/$T$)/d$T$ of the specific heat and the kink in the temperature derivative of the resistivity indicate the presence of a phase transition at about 0.45 - 0.5 K. With the local moment behavior observed in the susceptibility, the obvious guess is that this anomaly marks the onset of magnetic order. To gain further insight, we performed low-temperature AC susceptibility measurements. The temperature dependence of the real part of AC susceptibility ($\chi'$) under zero and various magnetic fields is presented in \figref{Fig2:res}(a). In zero-fields, $\chi'$ exhibit a pronounced discontinuity in its slope d$\chi'$/d$T$ at $T$$_{0}$ = 0.45 K. The change from a negative slope above $T$$_{0}$ to a positive one below $T$$_{0}$ is characteristic of antiferromagnetic (AFM) ordering. This AFM order is very sensitive to an external field. In the specific heat data, a field of 0.3 T is sufficient to smear out the sharp discontinuity in d($C$/$T$)/d$T$, resulting in a continuous evolution of the slope. The specific heat data do not provide a clear indication of whether this feature shifts to lower or higher temperatures at low fields. Only for a field larger than 0.5 T does it clearly shift to a higher temperature as the field increases. However, since at the same time the shape of the anomaly evolves towards a broadened Schottky-type anomaly, this feature can no longer be connected with an AFM transition. Instead, it corresponds to the doublet being Zeeman-split in the external field. This indicates that a field of the order of 0.5 T is sufficient to suppress the AFM order. Further insight into the low-field behavior was provided by the ac-susceptibility measurements in an external field (inset of \figref{Fig2:res}(a)). Although the anomaly becomes weaker and broadens with increasing magnetic field, making it more difficult to resolve, the results reveal a clear shift of $T$$_{0}$ to lower $T$, with $T$$_{0}$ $\approx$ 0.33 K at 0.15 T. This supports the magnetic order being a very weak AFM one. The observation that the magnetic order is very weak against an external field is in agreement with the observation that the magnetization curve $M$(H) at 1.8 K (inset of \figref{Fig1:ALL}(a)) is close to that expected for a single $S$ = $\frac{1}{2}$ ion \cite{S=1_2}. The discontinuities in the slope d$R$/d$T$, d($C$/$T$)/d$T$, and d$\chi'$(T)/d$T$ at H = 0 clearly indicate a phase transition, and the latter indicate it is of AFM type. However, the shape of the anomaly in $C$(T) does not correspond to either a mean-field type or a lambda type as expected for a second-order transition. Instead, a discontinuity in the derivative of the specific heat would correspond to a third-order type transition. However, the huge value of $C$/$T$ $\approx$ 2J/mol$_{Ce}$-K$^{2}$ just above $T$$_{0}$ and the small exponent n $\approx$ -0.9 observed in the $T$ dependence of $C$(T)/$T$ above $T$$_{0}$, which implies that the range with enhanced specific heat extends far above $T$$_{0}$, up to 10 K = 20 $T$$_{0}$, indicate that this AFM order occurs on the background of strong fluctuations of non-local nature, $i.e.$, of Kondo type. Therefore, the unusual type of anomaly is likely connected with weak AFM-type fluctuations competing with strong Kondo-type fluctuations. On the other hand, the sharpness of the discontinuities in d$C$(T)/d$T$ and d$\chi'$(T)/d$T$ excludes the absence of a lambda or mean-field anomaly to be due to disorder. Thus, analysis of the specific heat, susceptibility, and resistivity at low $T$ provides evidence of the onset of a weak AFM order at $T$$_{0}$ $\approx$ 0.45 K, formed on the background of strong Kondo-type fluctuations. The AFM order is so weak that it gets suppressed by a field of merely 0.5 T.

\begin{figure*}[t] 
\includegraphics[width=2.0\columnwidth]{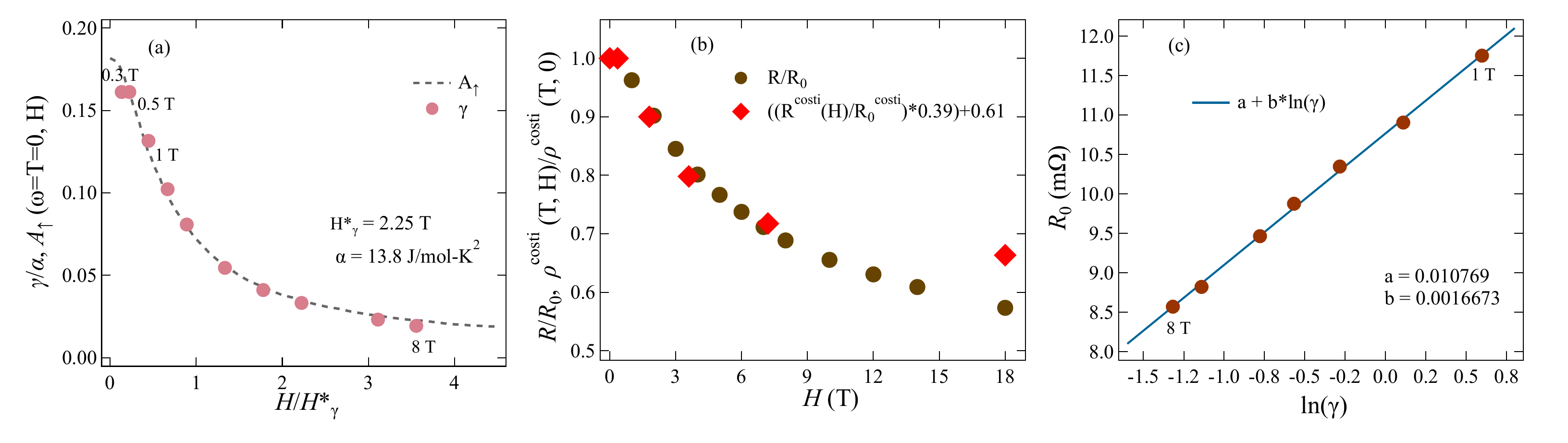} 
\caption{\label{Fig3:scaling} a) $\gamma$(H) scaling with $A$$_{\uparrow}$ which corresponds to the field-dependent spectral density per spin at $T$ = 0 at the Fermi level from Costi $et$. $al$. \cite{TA_Costi}. b) Field dependence of $R$/$R$$_{0}$ in the field range 1 T to 18 T scaled with Costi's theoretical normalized resistivity curve \cite{TA_Costi}. c) Scaling of $R$$_{0}$ as a function of ln($\gamma$).}
\end{figure*}

\subsection{Field-induced Fermi Liquid behavior}
Zero-field behavior of $C$(T)/$T$ and $R$(T) exhibits a marked departure from Fermi liquid theory, characterized by divergent behavior towards low $T$ in the first and non-observance of $T$$^{2}$ dependence down to low $T$ in the second. Application of a magnetic field results in two distinct regimes: i) In the low-field regime ($H$ $\leq$ 1 T), the magnetic transition is suppressed in both specific heat and resistivity measurements, 2) for larger fields ($H$ $\geq$ 1.5 T in $C$(T) and $H$ $\geq$ 2 T in $R$(T)), Fermi liquid behavior appears and gets fully established at the highest applied fields. This is best seen in the evolution of the resistance as a function of $T$ and $H$ (\figref{Fig2:res}(b)). At zero field, $R$(T) is nearly independent of $T$ for $T$ < 0.3 K, showing only a very weak decrease that is hard to discern due to data scatter. In contrast, already at $\mu_{0}$$H$ = 1 T, a clear decrease of $R$(T) with decreasing $T$ appears below 1.5 K, resulting in a maximum in $R$(T) at around 2 K. In Kondo systems, a maximum in $R$(T) at low T, followed by a pronounced decrease towards lower temperatures, is commonly attributed to the onset of a coherent Fermi-liquid state. Indeed, for larger fields $\mu_{0}$$H$ $\geq$ 2 T, $R$(T) evolves towards a Fermi liquid behavior $R$(T) = $R$$_{0}$ + $AT^{2}$ at lowest temperatures. This power law is well-established for $\mu_{0}$$H$ $\geq$ 3 T. The $T$ range where the quadratic dependence is observed increases significantly with increasing field, extending to above 5 K at the highest investigated field of 18 T. Simultaneously, the $A$ coefficient strongly decreases with increasing field (lower inset of \figref{Fig2:res}(c)).

A comparison with the evolution of the electronic part of the specific heat $C$$_{el}$(T)/$T$ as a function of $T$ and $H$ supports the interpretation of a field-induced Fermi liquid state. The raw data $C$(T)/$T$ versus $T$ presented in \figref{Fig1:ALL}(c) show the evolution from an almost $T$-independent $C$/$T$ behavior below $T$$_{0}$ = 0.45 K at zero field towards a more complex behavior at higher fields. There, one observes a maximum at temperatures above 1 K, which can be interpreted as the Schottky anomaly of the Zeeman-split ground-state doublet, a minimum in the range 0.2 K < $T$ < 0.5 K, and a large tail towards the lowest $T$. The latter can safely be attributed to the nuclear specific heat expected from the Ga- and Rh-nuclei. After subtraction of this nuclear part (see supplementary Section 4), a well-defined $T$-independent $C$$_{el}$/$T$ emerged, extending e.g. from 0.04 K up to 0.4 K at $\mu_{0}$$H$ = 7 T, over one $T$ decade. This $T$-independent $C_{el}$/$T$ can thus be interpreted as the electronic Sommerfeld coefficient $\gamma$ of a Fermi liquid state. This $\gamma$ value also decreases significantly with increasing magnetic field (upper inset of \figref{Fig2:res}(c)). Notably, the field dependence of $\gamma^{2}$ scales nicely with the field dependence of the $A$-coefficient of the resistivity, see \figref{Fig2:res}(c), as expected for a Fermi liquid. Therefore, the observation at the lowest $T$ and $H$ $\geq$ 2 T of a quadratic $T$ dependence in the resistivity, $R$(T) = $R$$_{0}$ + $AT^{2}$, a $T$-independent $C$$_{el}$(T)/$T$ = $\gamma$ value, and a scaling $A$ $\propto$ $\gamma$$^{2}$ provides conclusive evidence for a field-induced Fermi liquid state for fields larger than 1 T. The strong decreases of $\gamma$ and of $A$ with increasing field indicate a strong decrease of the effective mass of the quasiparticles. Such a decrease of the effective mass with increasing magnetic field is quite typical for Kondo lattices. It can be attributed to the suppression of the Kondo effect by competition with the field-induced Zeeman splitting. But in the present case, we also observe a pronounced decrease in the residual resistivity with increasing magnetic field, accompanied by a marked evolution from a fully incoherent regime at zero field to a well-defined Kondo lattice behavior at high field. Such behavior has rarely been observed; the most prominent example is UBe$_{13}$.

\subsection{Field dependence of Sommerfeld coefficient}

As stated above, the pronounced decrease in the effective mass of quasiparticles with increasing field is quite common in Kondo lattice systems, but the field scale at which it occurs varies widely. To gain further insight, we analyze $\gamma$ and its field dependence by comparing them with theoretical predictions. For a Kondo lattice system, the theoretical result would depend heavily on the details of the compounds and would require a specific calculation, which is far beyond the scope of the present paper. Therefore, we instead choose the single $S$ = $\frac{1}{2}$ Kondo impurity model, which exhibits universal behavior. Empirically, the specific heat of the Kondo lattice systems has been observed to be only weakly sensitive to lattice-induced coherence effects, in contrast to the resistivity. Since a detailed calculation of $\gamma$(H) can not be found in the literature, the spectral density $A$(H) calculated by T. A. Costi \cite{TA_Costi} was used, where $A$(H) stands for the density of states per spin at the Fermi level at $T$ = 0. $\gamma$(H) should be proportional to $A$(H).

Figure \ref{Fig3:scaling}(a) shows a very good agreement between the theoretical $A$(H) taken from Fig. 2 of \cite{TA_Costi} and the scaled experimental $\gamma$(H).  The latter was scaled vertically by a coefficient $\alpha$ = 13.8 J/mol-K$^{2}$/(states/T$_{K}$/spin), and horizontally by a Kondo Field $H$$^{*}_{\gamma}$ = 2.25 T.  Here, $H$/$T$$_{K}$ used in \cite{TA_Costi} was replaced by $H$/$H$$^{*}_{\gamma}$ for the sake of clarity because analysis of different properties commonly results in different values for the $T$$_{K}$ scale.

Both the vertical and horizontal scales can be used to determine the Kondo scale. For the vertical axis, we use the standard result $\gamma$ = $\frac{\pi}{3T_{K}}$ \cite{TA_Costi}. We checked that the definition of T$_{K}$ in \cite{TA_Costi} corresponds to the definition in \cite{TK}. Then $\gamma$ = 2.2 J/K$^2$mol$_{Ce}$ results in $T$$_{K,\gamma(0)}$ = 3.96 K. This value is close to, but slightly below, the value deduced from the analysis of the entropy.

The field scale provides further access to $T$$_{K}$. Using the standard relation $\mu_{B}$[T] = $\frac{\mu_{B}}{k_{B}}$ = 0.667 [K], the observed scaling field H$^{*}_{\gamma}$ = 2.25 T translate into $T$$_{K, H_{\gamma}}$ = 1.51 K. However, in \cite{TA_Costi}, the gyromagnetic factor was set to unity ($g$ = 1), as is usually done in theoretical calculations. In Ce systems, the effective $g$ factor of the CEF ground state is generally much larger \cite {NRef1, TK}. Notably, the saturation magnetization expected for an effective $S$ = $\frac{1}{2}$ system is $m$$_{sat}$ = $\frac{1}{2}$g [$\mu_{B}$]. Then $g_{eff}$ = 1 would correspond to $m$$_{sat}$ = 0.5 $\mu_{B}$. However, the magnetization shown in the inset of \figref{Fig1:ALL}(a) requires $m$$_{sat}$ $\geq$ 0.8 $\mu_{B}$, which implies $g$ $\geq$ 1.6. Correcting the deduced Kondo scale by a factor of 1.6 yields $T$$_{K, H_{\gamma}^{*}}$ = 2.42 K.

This is still significantly below $T$$_{K, \gamma(0)}$ = 3.96 K. Since for a Ce CEF ground state doublet, the saturation moment can be as high as 2.14 $\mu_{B}$, corresponding to a $g$$_{eff}$ = 4.29 in an $S$ = $\frac{1}{2}$ model, the difference could be attributed to a larger $g$ than suggested by the saturation moment. In order to recover the value $T$$_{K, \gamma(0)}$ = 3.96 K, a $g$ factor of 2.62, corresponding to a saturation moment of 1.31 $\mu_{B}$, is required. This is much larger than any direct estimation of $m$$_{sat}$, indicating that the real origin of this small field scale is an enhanced sensitivity of the quasiparticle effective mass to magnetic field. On a general level, our analysis of the field dependence of the specific heat shows that the field dependence of the density of states nicely follows the behavior predicted for a single Kondo ion, but the field scale is smaller than expected, $i.e.$, the sensitivity of heavy quasiparticles to an external field is stronger than expected.

\subsection{Field-induced decrease of the residual resistivity}

The prominent unusual feature in Ce$_{2}$Rh$_{3}$Ga$_{5}$ is the strong decrease of the residual resistivity $R$$_{0}$(H) = $R$(T$\to$0, H). In \figref{Fig3:scaling}(b), we plot $R$$_{0}$(H), as obtained from \figref{Fig2:res}(b), in function of $\mu_{0}$$H$. $R$$_{0}$ decreases by almost 40\% in the field range 0 $\leq$ $\mu_{0}$$H$ $\leq$ 18 T. Such a strong decrease is possible for a single Kondo impurity system, but not for a Kondo lattice; heavy quasiparticles are expected to form coherent Bloch states already at $H$ = 0. Large disorder may result in a large residual resistivity, as we observe at $H$ = 0, but the field dependence is expected to be very weak. In order to get more insight, we compare the observed $R$$_{0}$(H) dependence with that predicted for a single Kondo impurity system. We use the same calculation of T. A. Costi \cite{TA_Costi} as for the analysis of the specific heat, allowing for a direct comparison of the field dependence of $\gamma$(H) and $R$$_{0}$(H). We fitted the experimental normalized $r$$_{0}$(H) = $R$$_{0}$(H)/$R$$_{0}$(H = 0) with the function $r$$_{0}$(H) = $r$$_{0,\infty}$ + (1 - $r$$_{0,\infty}$)$\cross$r$_{0,cal}$(H), where $r$$_{0,\infty}$ corresponds to the static disorder induced residual resistivity part, assumed to be field independent, and $r$$_{0,cal}$(H) is the predicted field dependence for a single-ion Kondo impurity taken from Fig. 4(a) of \cite{TA_Costi}. In the latter, the field was scaled by a Kondo field $H^{*}_{r}$. We observed that it was not possible to obtain a reasonable fit across the entire field range. The low-field $\mu_{0}$$H$ $\leq$ 8 T regime can be fitted quite well, but above 8 T the experimental $R$$_{0}$(H) value decreases much faster than the theoretical prediction (\figref{Fig3:scaling}(b)). We note that this contrasts with the behavior of the Sommerfeld coefficient (\figref{Fig3:scaling}(a). There we do not have data above 8 T, but at 8 T the slope $\partial\gamma$/$\partial$H is already quite small. In the range $\mu_{0}$$H$ $\leq$ 8 T, a good fit could be obtained with $r$$_{0,\infty}$ = 0.61 and $H$$^{*}_{r}$ = 4.3 T. Notably, the field scale for the decrease in the residual resistivity is significantly larger than the field scale for the decrease in the Sommerfeld coefficient, almost a factor of 2. The much larger field scale for the suppression of the residual resistivity, as well as the difference in the high-field behavior, indicates that the field dependence of these two quantities cannot be fully accounted for by a single, identical mechanism.

However, in the region 0 $\leq$ $\mu_{0}$$H$ $\leq$ 8 T, we surprisingly found a simple relation between both quantities. \figref{Fig3:scaling}(c) demonstrates that $R$$_{0}$(H) is an almost perfect linear function of the logarithm of the Sommerfeld coefficient ln($\gamma$).  This is unexpected and somewhat surprising, because both the Sommerfeld coefficient and the resistivity are broadened measures of the total spectral density; one would expect a direct proportionality, not a logarithmic dependence. We do not know the origin of the observed logarithmic dependence, but it might be taken as evidence that the field dependence of the residual resistivity is not governed by the simple Kondo effect, but rather by an additional mechanism.

\begin{figure}[h] 
\includegraphics[width=1.0\columnwidth]{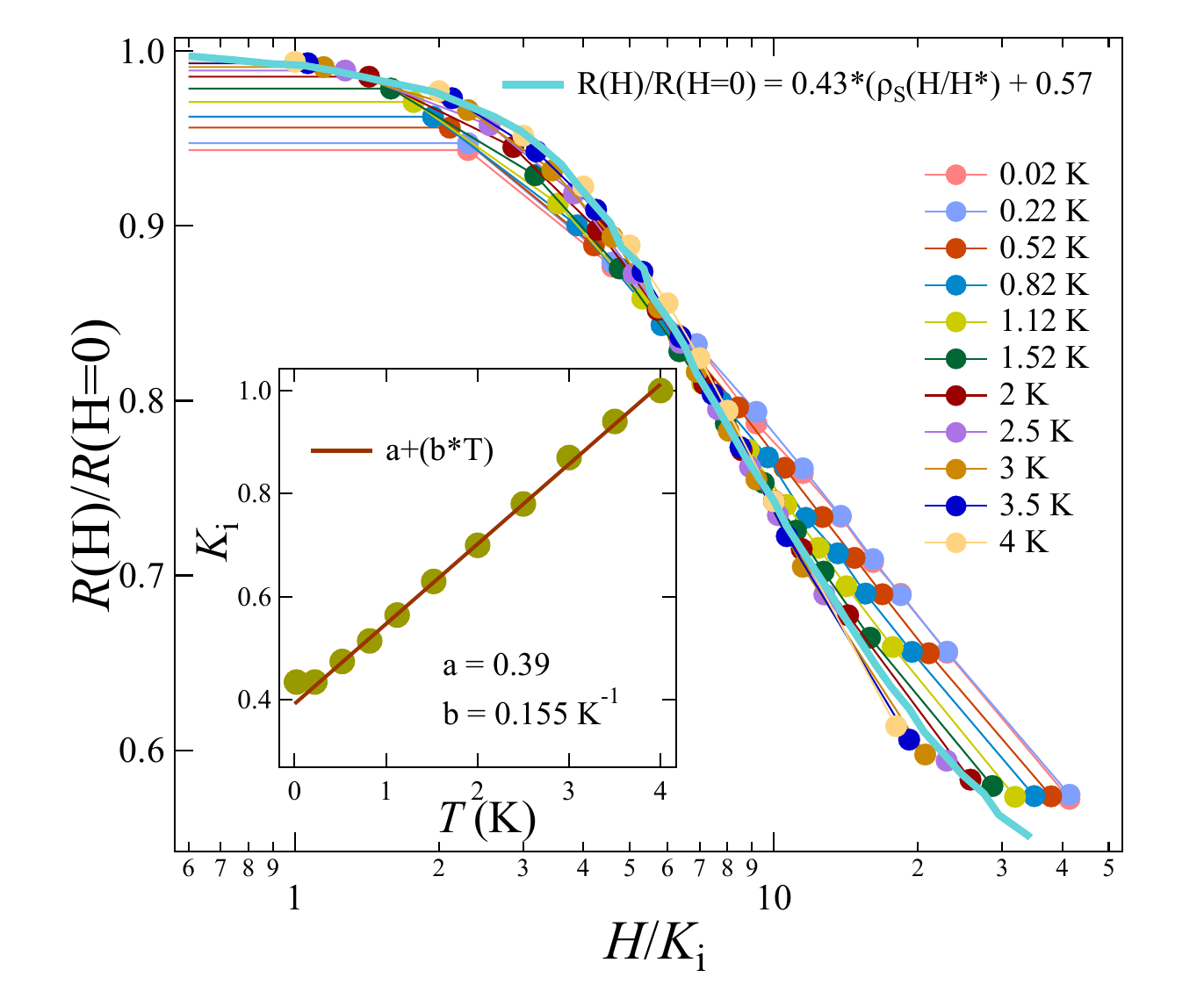} 
\caption{\label{Fig4:MRS} Absolute magnetoresistance isotherms (0.02 K - 4 K) shown on a scaled field axes $H$/$K_{i}$ with theoretical scaled Schlottmann curve for $S$ = $\frac{1}{2}$ Kondo impurity limit \cite{Schlottmann}. The inset shows the variation of the Kondo field $K_{i}$ with temperature.}
\end{figure}

For a further check of the extent to which the field and temperature dependence of the resistivity can be ascribed to a single mechanism, e.g., Kondo scattering, or requires a combination of different mechanisms, we tried to scale the T and H dependence of $\rho$(H)/$\rho$(T) using the standard approach used in Kondo systems: $R$(H, T)/$R$(H = 0, T) = f($\frac{H}{K}$). There, f($\frac{H}{K}$) is a universal function, depending on the properties of the local moment, and $K$ is a $T$-dependent field scaling factor. $K$ is generally assumed to be linear in $T$, $K$ = k$\cross$(T$_{K}$ + $T$), where $T$$_{K}$ is the Kondo scale and $k$ is just the ratio between thermodynamic and magnetic energy scales, $k$ = $\frac{k_{B}}{m}$, ($k_{B}$: Boltzmann constant, $m$: 4$f$ moment). For practical reasons and in order to be independent of a theoretical prediction for f($\frac{H}{K}$), we choose the curve for the highest investigated temperature, $T$ = 4 K, as a reference, and scale all other curves for further temperatures on this reference using an empirical $K_{i}$ (\figref{Fig4:MRS}). The reason for choosing the $T$ = 4 K isotherm as the baseline was that we suspect intersite interactions and coherence should have the least effect at high temperatures. We were only able to obtain an approximate scaling because the high and low temperature curves have slightly different dependencies on the scaled field, with the high temperature curve being steeper than the low temperature ones. Therefore, we choose $K_{i}$ so that all curves meet at the midpoint of the studied range, $i.e.$, at $\frac{H}{K_{i}}$ = 0.8. There is approximate scaling behavior, but systematic differences between high and low temperatures are clearly evident. This suggests that the behavior in high- and low-temperature regimes is dominated by two distinct mechanisms. Nevertheless, we analyzed the $T$ dependence of the scaling factor $K_{i}$. We found that $K$$_{i}$ follows a linear dependence, $K_{i}$ = $a$ + $b$$\cross$K$_{i}$, as usually assumed for the single-ion case; see the inset of \figref{Fig4:MRS}. From $a$ = 0.39 and $b$ = 0.155 K$^{-1}$, we deduce $T$$_{K}$ = 2.5 K. This $T$$_{K}$ corresponds to the field scale at $T$ = 0, but, as we shall discuss below, a comparison with other results requires additional input and analysis.

To gain further insight, we included a comparison with the theoretical result for the single-impurity Kondo model. Here, we used Schlottmann's prediction \cite{Schlottmann}, as it provides a complete curve, whereas Costi \cite{TA_Costi} provides only a few fixed points. As for the residual resistivity, we found that, for a reasonable fit, we need to account for an intrinsic, field-independent residual resistivity arising from static defects. The fitted theoretical curve was thus $R$(H)/$R$(H = 0) = 0.43 $\rho$$_{S}$(H/H$^{*}$) + 0.57, where $\rho_{S}$(H) is the result predicted by Schlottmann, $H$$^{*}$ is the field scale, and the factor 0.57 accounts for the field-independent contribution. This function fits the high temperature curve, while the low $T$ curve deviates significantly (\figref{Fig4:MRS}). This suggests that the high temperature curve is dominated by the single-impurity Kondo effect, as we initially expected, while an additional mechanism is active at low $T$. This comparison provides access to an absolute value for the field scales $H$$_{i}$. At $T$ = 4 K, $H$$_{i}$ = 17.8 T. This translates into parameters $a$ = 6.9 T and $b$ = 2.72 T/K. The parameter $a$ corresponds to the field scale at $T$ = 0, $H$$^{*}_{scaling}$ = 6.9 T, which is even larger than the field scale for the suppression of the residual resistivity $H$$^{*}_{r}$ = 4.3 T, and much larger than the field scale for the suppression of the Sommerfeld coefficient $H$$^{*}_{\gamma}$ = 2.25 T. Using the standard ratio between magnetic and thermal energy (see above), one can infer from parameter $b$ a saturation moment of only 0.544 $\mu_{B}$, in agreement with the observation that the field scales are much larger than the temperature scales.  Thus, this analysis confirms that the field scale for suppressing resistivity is much larger than that for suppressing the Sommerfeld coefficient. This large difference, and the difference in the field dependence of the resistivity at high and low temperatures, prevent a general $T$-$H$ scaling, indicating that the field dependence of the resistivity is governed not by a single mechanism but by at least two distinct mechanisms. The observation that the scaling with the prediction for a single Kondo ion is much better at high temperature than at low temperature suggests that the Kondo effect is the dominant mechanism at high temperatures similar to UBe$_{13}$ \cite{UBe13_highT1,UBe13_highT2}. The nature of the dominant mechanism at low $T$ is difficult to access. A likely candidate is weak localization. However, we are not aware of a precise prediction for the combined effect of Kondo and weak localization on the resistivity in a Kondo lattice system.  

\section{Summary}

In summary, we investigated the unusual thermodynamic and magnetotransport properties of the newly discovered Ce-based heavy fermion compound, Ce$_{2}$Rh$_{3}$Ga$_{5}$, which exhibits striking similarities to the anomalous normal state behavior of UBe$_{13}$. The $T$ dependence of the resistivity with 2 maxima at around 100 K and 1 K, and the slow increase of $C$/$T$ below 4 K, ending in a value of 2.2 J/mol$_{Ce}$K$^{2}$ below 0.4 K, indicate a sizeable Kondo interaction resulting in the formation of heavy quasiparticles at low $T$. However, $\rho$(T) does not show the decrease expected for the formation of a coherent Kondo lattice; instead, it remains at a very large value down to the lowest $T$. Anomalies in the specific heat, the ac-susceptibility, and the resistivity at $T$$_{0}$ = 0.4 K, and the disappearance of these anomalies in a field $\mu_{0}$$H$ < 0.5 T indicate the transition to a very weak AFM state below $T$$_{0}$. Applying a field larger than 1 T induces a well-defined Fermi-liquid state in $\rho$(T, H) and $C$(T,H)/$T$, with a strong reduction in the effective mass. Its field dependence can be nicely described within the single-ion Kondo model, but with a smaller field scale than expected from $T$$_{K}$ deduced from the $H$ = 0 thermodynamic properties. The surprising result is the observation of a strong decrease of the residual resistivity with increasing magnetic field, by nearly 40\% at 18 T, where $\rho$(H) still does not show evidence for saturation. This indicates that the huge residual resistivity observed at $H$ = 0 cannot be attributed to static disorder. In contrast to $C$/$T$, neither the residual resistivity nor the magnetoresistance could be well-fitted within the single-ion Kondo model. Furthermore, an attempt to scale our extensive $\rho$(T, H) magnetoresistance data failed, revealing distinct field dependencies at low and high fields and a field scale much larger than that observed for $C$(T,H)/$T$. These observations suggest that the magnetoresistance involves at least two distinct scattering contributions, with conventional single-ion Kondo scattering dominating at elevated temperatures and an additional mechanism becoming important at low temperatures. Although the microscopic origin of the latter remains unresolved, weak localization represents a possible candidate. The striking resemblance between Ce$_{2}$Rh$_{3}$Ga$_{5}$ and UBe$_{13}$ suggests that such anomalous magnetotransport may reflect more general interactions in strongly correlated electron systems rather than a compound-specific phenomenon. Importantly, unlike UBe$_{13}$, the absence of superconductivity in Ce$_{2}$Rh$_{3}$Ga$_{5}$ enables investigation of these anomalous scattering processes down to the lowest temperatures across the accessible magnetic field range.

\section{Acknowledgments}

We thank S. Kostmann and P. Scheppan for EDXS measurements, as well as S. Scharsach and M. Schmidt for performing DTA experiment.

\clearpage

\title{Supplementary Information to "Anomalous Field-induced Coherence beyond Single-ion Kondo Physics"}\label{ExtractParameter}
\maketitle

\subsection{Experimental Details}\label{}
The Ce$_{2}$Rh$_{3}$Ga$_{5}$ sample was first prepared using the standard arc-melting technique. The constituent elements Ce (99.99\%), Rh(99.95\%), and Ga(99.999\%) were taken in stoichiometric ratios and melted together in the arc furnace under an argon atmosphere. The resulting ingot was flipped over, remelted, and this process was repeated three times to promote homogeneity. It resulted in an almost phase-pure polycrystalline sample of Ce$_{2}$Rh$_{3}$Ga$_{5}$ with 2-3\% of another phase, which might have formed due to the evaporation of gallium. A 40 mg sample from this batch was used to perform a DTA measurement, which suggested that Ce$_{2}$Rh$_{3}$Ga$_{5}$ forms almost congruently and that the melting temperature is around 1100 $\degree$C. For the single-crystal synthesis, a pre-reaction of the constituent elements Ce, Rh, and Ga (3\% excess relative to the stoichiometric composition to compensate for evaporation losses) was first performed in the arc furnace. Small crushed pieces of the pre-reacted ingot were put in a graphite crucible, which was then enclosed in a tantalum cylindrical crucible. The tantalum crucible was sealed using the arc furnace under an argon atmosphere of 800 mbar. The sealed tantalum crucible was then placed in an electric resistance furnace, which was ramped above the melting temperature (1100 $\degree$C), held there for some time, and then slowly cooled to room temperature. This synthesis produced phase pure single crystals of Ce$_{2}$Rh$_{3}$Ga$_{5}$. To check the phase purity and the crystal structure of both polycrystalline samples and single crystals, powder X-ray diffraction (XRD) patterns were obtained using a Stoe-Stadi-MP powder diffractometer in transmission mode equipped with CuK$_{\alpha1}$ radiation ($\lambda$ = 1.54056 \AA), curved germanium (111) Johann-type monochromator, and DECTRIS MYTHEN2 1K silicon strip detector. Phase purity and composition were also assessed by energy-dispersive X-ray analysis (EDX) of polished sample pieces. Magnetization measurements in the temperature range 1.8 K - 300 K and in applied fields up to 7 T were done on the Quantum Design MPMS3 magnetometer. Specific heat measurements using the two-tau relaxation method were performed on the Quantum Design PPMS equipped with a $^{3}$He probe down to 0.5 K. A compensated heat-pulse calorimeter was used to measure specific heat from 25 mK to 4 K in the dilution refrigerator. Electrical resistivity in four-probe geometry was first measured with a Quantum Design PPMS equipped with a $^{3}$He probe down to 0.5 K. Measurements of electrical resistivity in the 0.02 - 5 K temperature range and in magnetic fields of up to 18 T were conducted with a Kelvinox K400 dilution refrigerator produced by Oxford Instruments. For a four-point resistivity measurement, gold wires were contacted to an approximately millimetre-sized irregularly-shaped sample with DuPont 4929 silver paste. The measurement was conducted with a lock-in amplifier (Stanford Research Systems SR830), using the excitation current of 30 $\mu$A (RMS), the frequency of 37 Hz, and the integration time of 1 second.

Measurements of AC magnetic susceptibility in the sub-Kelvin temperature range were conducted in a Kelvinox MX400 dilution refrigerator produced by Oxford Instruments. The cryostat was equipped with a superconducting modulation coil which was used for generating the driving AC magnetic field. The RMS amplitude of the modulation field was in the 0.18 -- 0.35 mT range. A crystal of Ce$_{2}$Rh$_{3}$Ga$_{5}$ of about 1 mm in size was placed inside a pick-up coil of a home-made susceptometer, which also included a counter-wound compensation coil. In order to ensure a good thermalization of the sample, a few cm long silver wire of the 50 $\mu$m diameter was attached to it with a silver paste. The wire was then connected to a longer 370 $\mu$m diameter silver wire, the opposite end of which was clamped to the mixing chamber plate of the cryostat. The measurement was conducted using a lock-in technique (implemented via a National Instruments PXI system) at the frequency of 5 Hz and with 1 second integration time. A measurement done with an empty coil allowed to eliminate the background signal due to an imperfect compensation of the susceptometer. The filling factor of the sample inside the coil has a large uncertainty (roughly estimated to be between 5 and 15\%). The presented data correspond to the real part of susceptibility up to a multiplicative factor. During the measurement, the superconducting magnet was found to have a -0.015 T remnant field. The values of the applied magnetic field provided in Fig. 2a include a correction due to the remnant field.

\subsection{Structural Characterization}\label{XRD1}
Figure \ref{Fig1:xrd} shows the Rietveld refinement of the powder XRD pattern collected at room temperature. To the best of our knowledge, there is only one report on the compound Ce$_{2}$Rh$_{3}$Ga$_{5}$ in the literature \cite{Ce2Rh3Ga5_ref1}, which provides only the structure type and no information on additional structural parameters. Therefore, to obtain the crystal structure parameters, we have used the crystallographic information file (CIF) of our Ce$_{2}$Ir$_{3}$Ga$_{5}$ compound and replaced Ir with Rh. The Rietveld refinement using this modified CIF confirms the phase purity and the crystal structure as Orthorhombic U$_{2}$Co$_{3}$Si$_{5}$ type with space group $Ibam$ (No. 72). All the parameters obtained from the refinement are tabulated in Table \ref{STP}. The change in the lattice parameters and cell volume of Ce$_{2}$Rh$_{3}$Ga$_{5}$ in comparison to Ce$_{2}$Ir$_{3}$Ga$_{5}$ is in agreement with replacing a larger atomic-radius atom (Ir) with a smaller atomic-radius atom (Rh).

\setcounter{table}{0}
\renewcommand{\thetable}{S\arabic{table}}
\begin{table}[H]
\caption{Crystal structure parameters obtained from Rietveld refinement of Ce$_{2}$Rh$_{3}$Ga$_{5}$ powder XRD.}
\label{STP}
\setlength{\tabcolsep}{18pt}
\begin{tabular}{l r} \hline\hline
Crystal Structure& Orthorhombic\\[0.5ex]
Space group, Z&	$Ibam$ (No. 72), 4\\[0.5ex]
Lattice parameters (\text{\AA})& a = 10.1235\\[0.5ex]
& 	b = 12.7764\\[0.5ex]
&  c =   5.7367\\[0.5ex]
V (10$^{6}$ pm$^{3}$)&	741.9918\\[0.5ex] \hline

\end{tabular}

\begingroup
\setlength{\tabcolsep}{7pt}
\begin{tabular}[b]{c c c c c c}
Atom&  Wyckoff position& x/a& y/b& z/c\\[1ex]
\hline
Ce1& 8j& 0.27525& 0.36057& 0\\
Rh1& 8j& 0.09354& 0.14471& 0\\
Rh2& 4b& 1/2& 0& 1/4\\                         
Ga1& 8j& 0.32819& 0.08683& 0\\
Ga2& 8g& 0& 0.29674& 1/4\\
Ga3& 4a& 0& 0& 1/4\\
[1ex]
\hline
\end{tabular}
\par\medskip\footnotesize
\endgroup
\end{table}

\setcounter{figure}{0}
\renewcommand{\thefigure}{S\arabic{figure}}
\begin{figure}[t]
\includegraphics[width=1.0\columnwidth,origin=b]{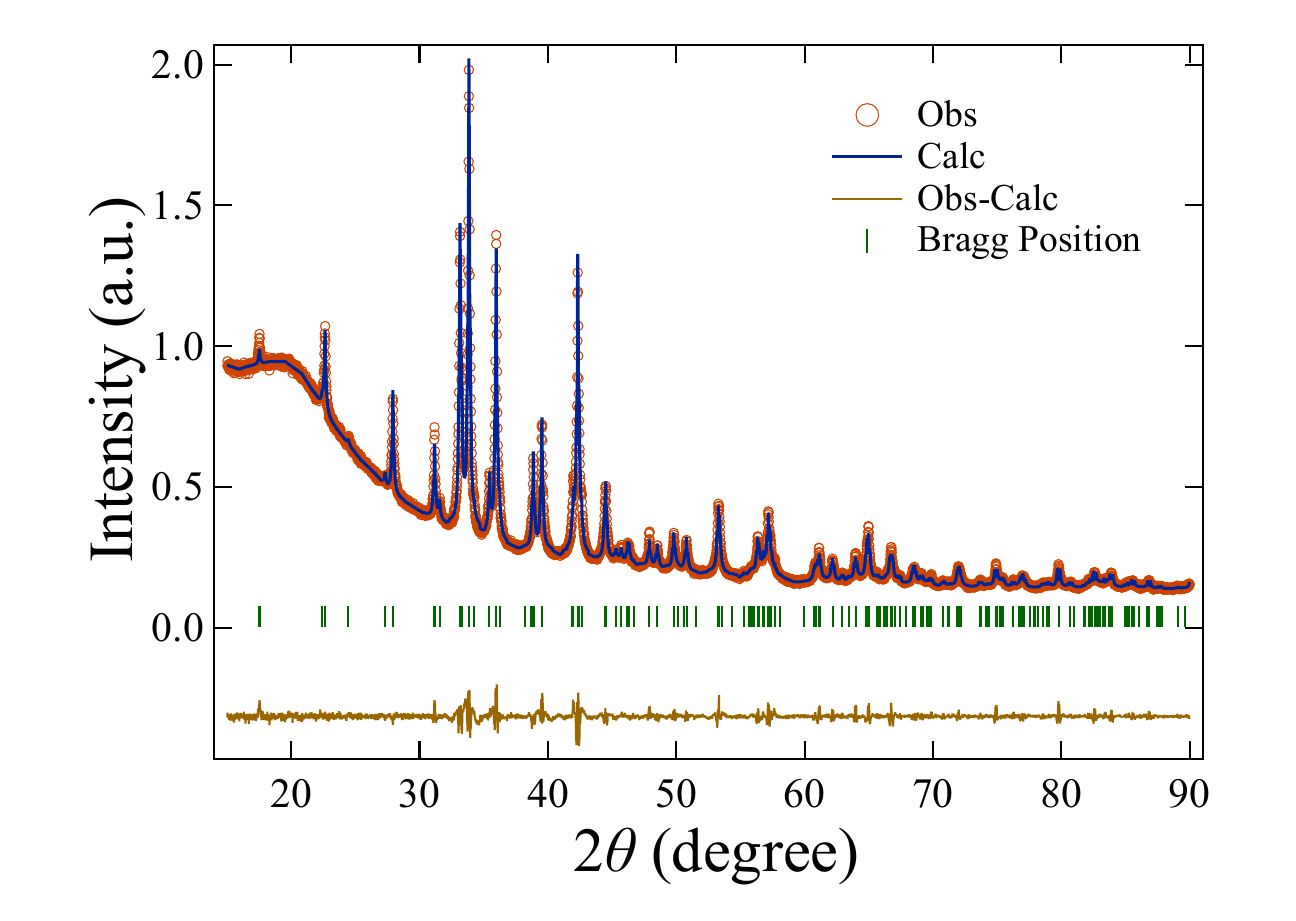}
\caption{\label{Fig1:xrd} Powder XRD pattern of Ce$_{2}$Rh$_{3}$Ga$_{5}$ and the Rietveld refinement are shown by red open circles and blue solid line, respectively. Small green bars indicate the Bragg reflection positions, and the solid brown line (bottom) shows the difference between the calculated and observed intensities.}
\end{figure}

\subsection{Magnetization}\label{Mag}
\begin{figure}[t]
\includegraphics[width=1.0\columnwidth,origin=b]{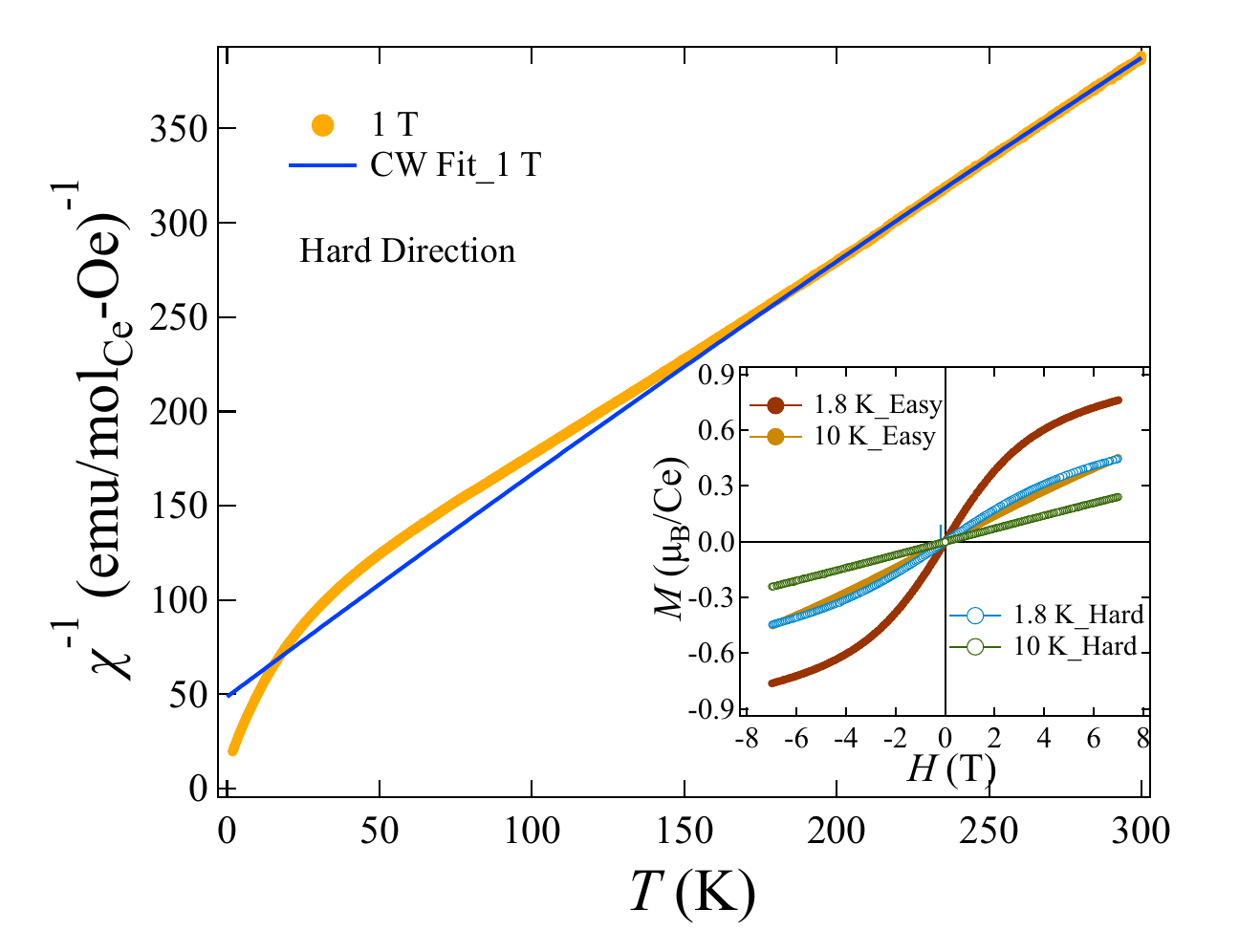}
\caption{\label{Fig2:mag} Temperature dependence of $\chi^{-1}$ in 1 T applied field for hard direction. The high-temperature part is well fitted with CW behavior. The inset represents M(H) curves taken at different temperatures, 1.8 K and 10 K, for the field along both hard and easy orientations.}
\end{figure}

Figure \ref{Fig2:mag} shows the temperature dependence of the inverse susceptibility $\chi^{-1}$(T) along the hard orientation in an applied field of 1 T. In the high-temperature regime $T$ > 150 K, $\chi^{-1}$(T) is well described by the modified Curie-Weiss (CW) equation: $\chi^{-1} (T) = \left(\chi_{0} + {C}/{(T-\theta_{p}})\right)^{-1}$, where $\chi_{0}$ is the temperature-independent Pauli susceptibility, $C$ is the Curie constant, and $\theta_{p}$ is the paramagnetic Curie constant. The fit provides $\chi_{0}$ = 1.86(1)$\cross10^{-4}$ emu/mol$_{Ce}$, $C$ = 0.81(1) emu-K/mol$_{Ce}$, and $\theta_{p}$ = -40(1) K. The obtained effective magnetic moment $\mu_{eff}$ = 2.55 $\mu_{B}$/Ce is close to the expected value (2.54 $\mu_{B}$) for a Ce$^{3+}$ free ion, indicating that Ce$_{2}$Rh$_{3}$Ga$_{5}$ is in a localized moment regime at high temperatures. In comparison to $\chi^{-1}$(T) at 1 T along the easy orientation (shown in Fig.1 in the main paper), there is a marked deviation from linearity for $T$ < 100 K. This is quite typical for a hard direction and can be attributed to the thermal depopulation of the excited crystal-electric field (CEF) levels. Similar to $\chi^{-1}$(T) along the first orientation, there was no anomaly observed down to the lowest measured temperature, suggesting that Ce$_{2}$Rh$_{3}$Ga$_{5}$ remains in the non-ordered state down to 1.8 K. The inset presents $M$(H) isotherms collected at 1.8 K and 10 K for fields along both hard and easy orientations. The $M$(H) isotherm at 1.8 K exhibits non-linearity similar to that in the easy direction. The estimated saturation moment for the hard axis direction is 0.44 $\mu_{B}$/Ce, which is about a factor of 2 smaller than along the easy direction. At 10 K, a linear magnetization response to the applied field is recovered for both the hard and easy axes. 

\subsection{Specific Heat}\label{SH}

\begin{figure}[t]
\includegraphics[width=1.0\columnwidth,origin=b]{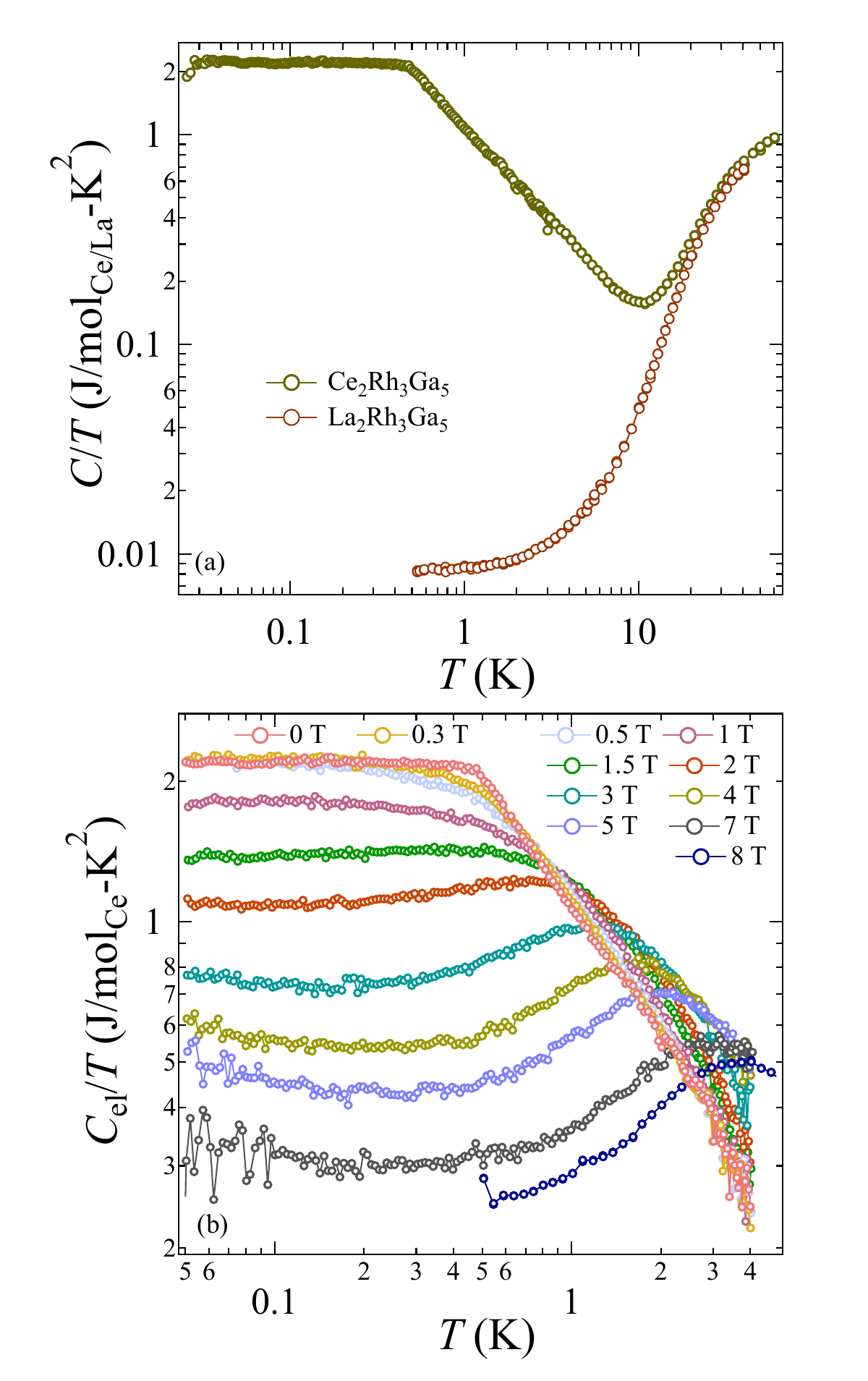}
\caption{\label{Fig3:SH1} Temperature dependence of $C$/$T$ in zero field for Ce$_{2}$Rh$_{3}$Ga$_{5}$ and La$_{2}$Rh$_{3}$Ga$_{5}$. Low temperature variation of electronic specific heat of Ce$_{2}$Rh$_{3}$Ga$_{5}$ in zero and applied fields up to 8 T.}
\end{figure}

Figure \ref{Fig3:SH1}(a) display the $T$ dependence of the specific heat of Ce$_{2}$Rh$_{3}$Ga$_{5}$ and of its non-magnetic homolog La$_{2}$Rh$_{3}$Ga$_{5}$ in a plot ln(C/T) versus ln(T). Both curves are very similar above 15 K, but diverges completely to lower temperatures. This reflects the difference between a classical standard metal and a heavy Fermion system. Thus $C$/$T$ of La$_{2}$Rh$_{3}$Ga$_{5}$ shows a very small, almost constant value below 2 K, as expected for the electronic contribution of a standard metal, and a strong increase for $T$ > 2 K, as expected for the phononic contribution. Extrapolation of the low-temperature data to 0 K yielded the Sommerfeld coefficient $\gamma$ = 8.4 mJ(mol$_{La}$)$^{-1}$K$^{-2}$, confirming that La$_{2}$Rh$_{3}$Ga$_{5}$ is a weakly correlated metal with no phase transition down to the lowest measured temperature of 0.5 K. On the other hand, $C$(T)/$T$ of Ce$_{2}$Rh$_{3}$Ga$_{5}$ is only slightly larger than $C$(T)/$T$ of La$_{2}$Rh$_{3}$Ga$_{5}$ for $T$ > 15 K, because there the specific heat is dominated by the phononic contribution, with the additional magnetic contribution of the Ce compound being much smaller (see also analysis of the entropy of Ce$_{2}$Rh$_{3}$Ga$_{5}$ in the main paper). Below 10 K, $C$(T)/$T$ of Ce$_{2}$Rh$_{3}$Ga$_{5}$ exhibits an upturn, leveling off to a constant value below 0.5 K. This increase differs from the steep rise characteristic of conventional magnetic systems near the magnetic ordering temperature. Instead, it follows a power law $C$(T)/$T$ $\propto$ T$^{-0.9}$, which is typical for heavy fermion systems close to a quantum critical point. Further features in $C$(T)/$T$ have been discussed in the main paper. For fields larger 1 T, the raw $C$(T)/$T$ curves in Fig. 1(c) show a pronounced divergence to very low T. This can safely be attributed to the nuclear contribution of Rh and Ga. Furthermore, the raw data also include the phononic contribution. In order to extract the magnetic contribution, we first fitted the low $T$ data with $C$(T)/$T$ = $\gamma$ + $\alpha$/T$^{3}$, where $\gamma$ and $\alpha$ represent the Sommerfeld and nuclear Schottky coefficients, respectively. The obtained nuclear contribution ($\alpha$/T$^{3}$) and the phononic contribution from the La analog were subtracted from the raw $C$(T) data to evaluate the magnetic contribution, $C$$_{el}$. The temperature dependence of $C$$_{el}$ is shown in \figref{Fig3:SH1}(b) in a plot ln($C$$_{el}$(T)/T) versus ln(T) for fields 0 < $\mu_{0}$$H$ < 8 T.

\begin{figure}[b]
\includegraphics[width=1.0\columnwidth,origin=b]{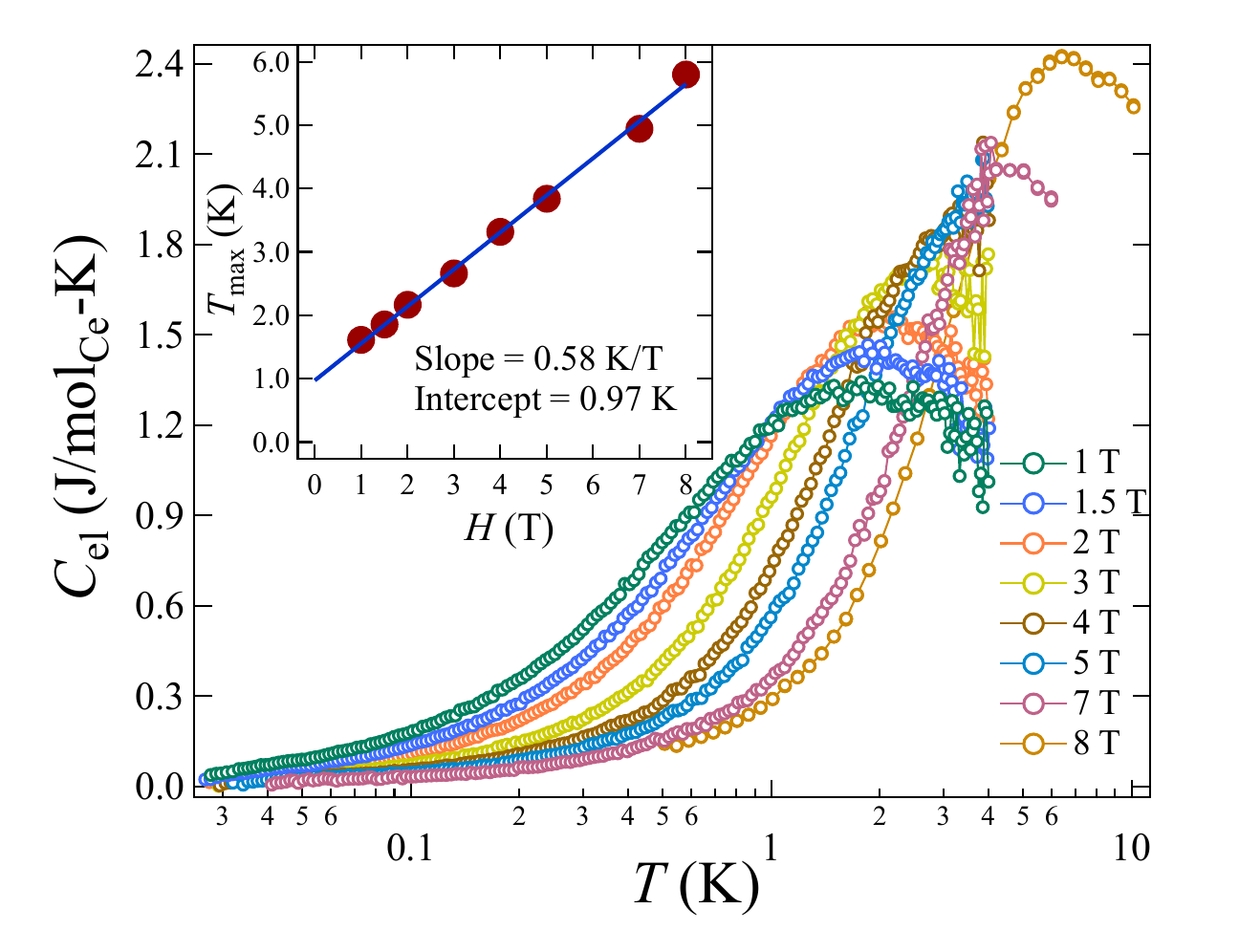}
\caption{\label{Fig4:SH2} Temperature evolution of magnetic specific heat, C$_{mag}$ of Ce$_{2}$Rh$_{3}$Ga$_{5}$ on a logarithmic temperature scale in external fields from 1 T to 8 T. Plot of T$_{max}$ $vs.$ external fields $H$.}
\end{figure}

For $H$ $\geq$ 1 T a maximum starts to form in $C$(T)/$T$ at T > 1 K, which becomes more pronounced, gets sharper, and shifts to higher $T$ with increasing field. To get more insight in this feature we plot the $T$ dependence of the specific heat in a plot $C$ versus ln(T) (Fig. S4). There it becomes clear that with increasing field this feature gets closer and closer to a Schottky anomaly. Therefore, this maximum can be interpreted as corresponding to the Zeeman splitting of the 4$f$ CEF ground state doublet in an effective magnetic field that is a combination of the external ($H$$_{ext}$) and internal ($H$$_{in}$) magnetic fields. The internal field is due to intersite exchange interactions and polarization of the local moments. The magnetic field dependence of the temperature at which the broadened maxima occur (denoted as $T$$_{max}$) is shown in the inset of \figref{Fig4:SH2}. 

Since at low temperature the Ce-4$f$ moment is an effective $J$ = 1/2 doublet, $T$$_{max}$ is given by $T$$_{max}$k$_{B}$ = 0.417 $\Delta$, where $\Delta$ is the Zeeman splitting of the doublet, $\Delta$ = 2m$_{sat}$$\mu_{B}$H$_{eff}$. As a result, $T$$_{max}$ is proportional to $H$$_{eff}$: $T$$_{max}$ = $\frac{\mu_{B}}{k_{B}}$m$_{sat}$H$_{eff}$, i.e. $T$$_{max}$ = 0.56m$_{sat}H_{eff}$  with $T$$_{max}$ in [K], m$_{sat}$ in $\mu_{B}$, and $H$ in [T]. When $H$$_{ext}$ $\gg$ $H$$_{in}$, the increase in $H$$_{eff}$ is equal to the increase in $H$$_{ext}$, but for $H$$_{ext}$ $\leq$ H$_{in}$, the relation is more complex and depends on the details of the intersite interactions. Indeed, the plot of $T$$_{max}$ versus $H$$_{ext}$ in the inset of \figref{Fig4:SH2} show a linear dependence on field for the whole investigated range 1 T $\leq$ $\mu_{0}$H$_{ext}$ $\leq$ 8 T, with the slope 0.58 K/T. From this slope one deduces a saturation moment m$_{sat}$ = 1.04 $\mu_{B}$, which is slightly larger than the value estimated from the $M$(B) curve at 1.8 K (Inset in Fig. 1 (a)). The slightly larger m$_{sat}$ value corresponds to a larger slope d($T$$_{max}$)/d($H$$_{in}$). This larger slope might originate from weakening of the Kondo screening with increasing $H$$_{ext}$, which would result in an increase of m$_{sat}$ with field. A linear fit to the $T$$_{max}$($H$$_{ext}$) data extrapolate to a finite, positive $T$$_{max}$ at $H$$_{ext}$ = 0. This implies that intersite exchange fields enhance the effective field, which indicate that the sum of all intersite exchanges is slightly ferromagnetic. This is not in contradiction with an AFM ordered ground state, a very simple possibility would be e.g. strong ferromagnetic in-plane interaction combined with a weaker AFM inter-plane interaction.

\begin{figure}[h]
\includegraphics[width=1.0\columnwidth,origin=b]{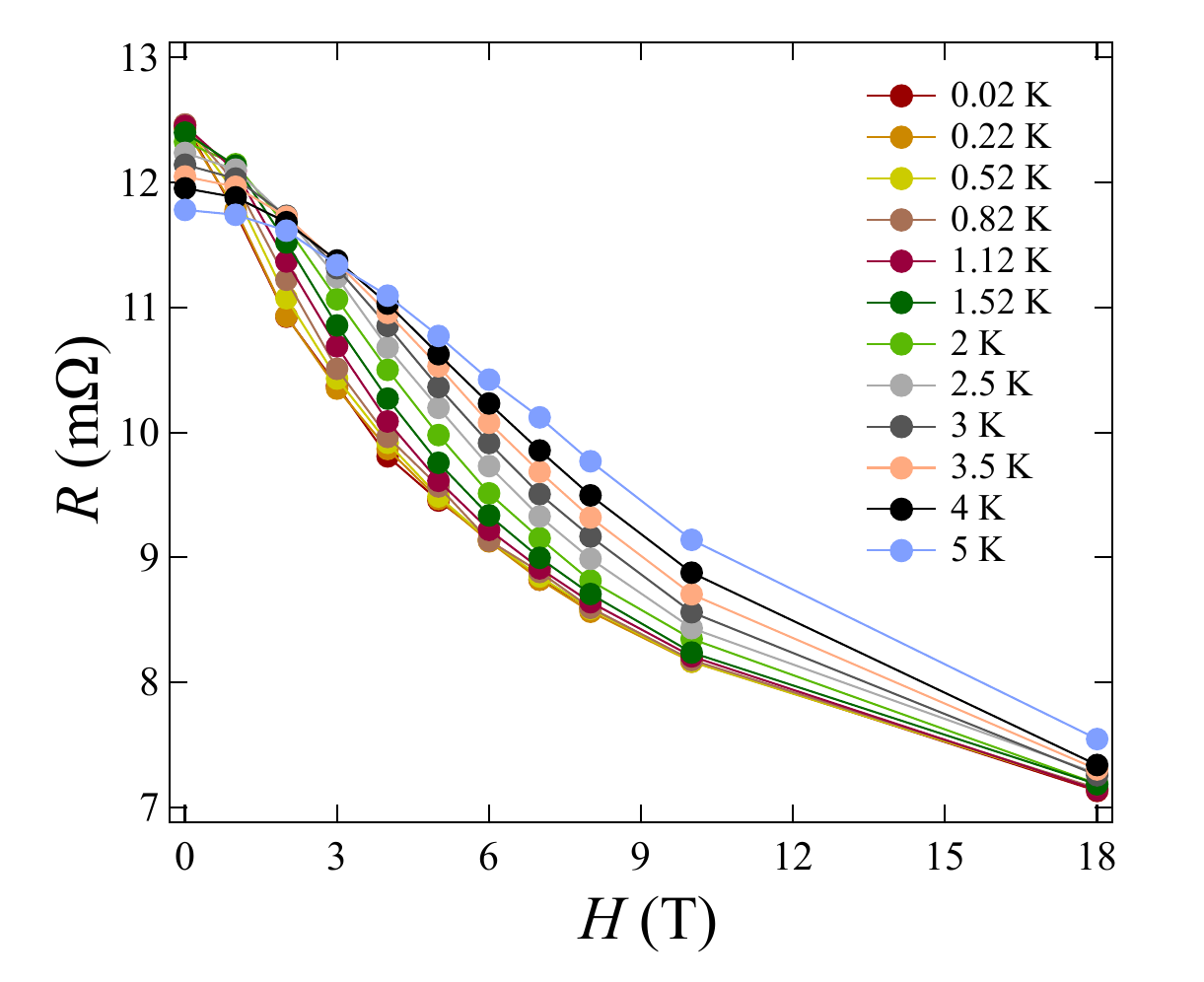}
\caption{\label{Fig5:rho_H} Field dependence of resistance at various temperatures, ranging from 0.02 K to 5 K.}
\end{figure}

\subsection{Electrical Resistivity}\label{SH}

The electrical resistivity was measured as a function of temperature at several fixed magnetic fields, as this approach is experimentally more efficient. However, analyzing the magnetoresistance requires the field dependence of resistivity at fixed temperatures. Accordingly, magnetoresistance isotherms, \(R(H,T=\mathrm{const.})\), were constructed from the measured \(R(T,H=\mathrm{const.})\) data over the temperature range 0.02-5 K. The resulting curves are presented in \figref{Fig5:rho_H}, with a detailed analysis and discussion provided in the main text.

\end{document}